\documentclass[conference]{IEEEtran}
\IEEEoverridecommandlockouts

\usepackage{cite}
\usepackage{amsmath,amssymb,amsfonts}
\usepackage{algorithmic}
\usepackage{graphicx}
\usepackage{textcomp}
\usepackage{xcolor}
\usepackage{hyperref}
\def\BibTeX{{\rm B\kern-.05em{\sc i\kern-.025em b}\kern-.08em
    T\kern-.1667em\lower.7ex\hbox{E}\kern-.125emX}}
\begin{document}

\title{Financial Fine-tuning a Large Time Series Model}

\author{
    \IEEEauthorblockN{Xinghong Fu}
    \IEEEauthorblockA{
        \textit{Massachusetts Institute of Technology} \\
        Cambridge, MA, United States \\
        fxh@mit.edu
    }
    \and
    \IEEEauthorblockN{Masanori Hirano}
    \IEEEauthorblockA{
        \textit{Preferred Networks, Inc.} \\
        Tokyo, Japan \\
        research@mhirano.jp
    }
    \and
    \IEEEauthorblockN{Kentaro Imajo}
    \IEEEauthorblockA{
        \textit{Preferred Networks, Inc.} \\
        Tokyo, Japan \\
        imos@preferred.jp
    }
}


\maketitle

\begin{abstract}
Large models have  shown unprecedented capabilities in natural language processing, image generation, and most recently, time series forecasting. This leads us to ask the question: treating market prices as a time series, can large models be used to predict the market? In this paper, we answer this by evaluating the performance of the latest time series foundation model TimesFM on price prediction. We find that due to the irregular nature of price data, directly applying TimesFM gives unsatisfactory results and propose to fine-tune TimeFM on financial data for the task of price prediction. This is done by continual pre-training of the latest time series foundation model TimesFM on price data containing 100 million time points, spanning a range of financial instruments spanning hourly and daily granularities. The fine-tuned model demonstrates higher price prediction accuracy than the baseline model. We conduct mock trading for our model in various financial markets and show that it outperforms various benchmarks in terms of returns, sharpe ratio, max drawdown and trading cost.  
\end{abstract}


\section{Introduction}
Predicting the market has long been of interest to researchers. Under the more general task of time-series forecasting, countless research attempts date back to simple moving averages \cite{mckenzie1984}, various types of models has been developed, including autoregressive \cite{box1970}, global univariate models like N-BEATS \cite{nbeats}, and long-term forecasting models \cite{patchtst}. 

Following the development of large language models \cite{Vaswani2017, bert, gpt}, researchers have also attempted to directly make use of the zero-shot forecasting capabilities of LLMs \cite{llmtime, timellm, llmtimeseriessurvey, llmp, sun2024testtextprototypealigned}. Benefits of using pre-trained LLMs include the availability of text context together with numeric data to improve forecasting prediction accuracy \cite{llmp}, and access to a strong encoder/decoder such that tuning the LLM outputs to time-series forecasting only requires aligning the embedding layer for numeric time series data \cite{sun2024testtextprototypealigned, timellm}. However, recent work \cite{tan2024languagemodelsactuallyuseful} raises question on the usefulness of the LLM backbone by comparing it to basic attention layers trained from scratch. 
A similar concern turns our attention to TimesFM\cite{timesfm}, a foundation time-series model trained from scratch, specifically for the task of time-series forecasting. Detailed more in Section \ref{relatedwork}, TimesFM achieves state-of-the-art performance of multiple forecasting benchmarks. However, these benchmarks most oftenly include regular and seasonal data, much unlike financial data that we are interested in. In this work, we answer the following research question: \textbf{can a foundation time series model perform well on price data in the financial markets?} 

\begin{figure}[t]
    \centering
    \includegraphics[width=\linewidth]{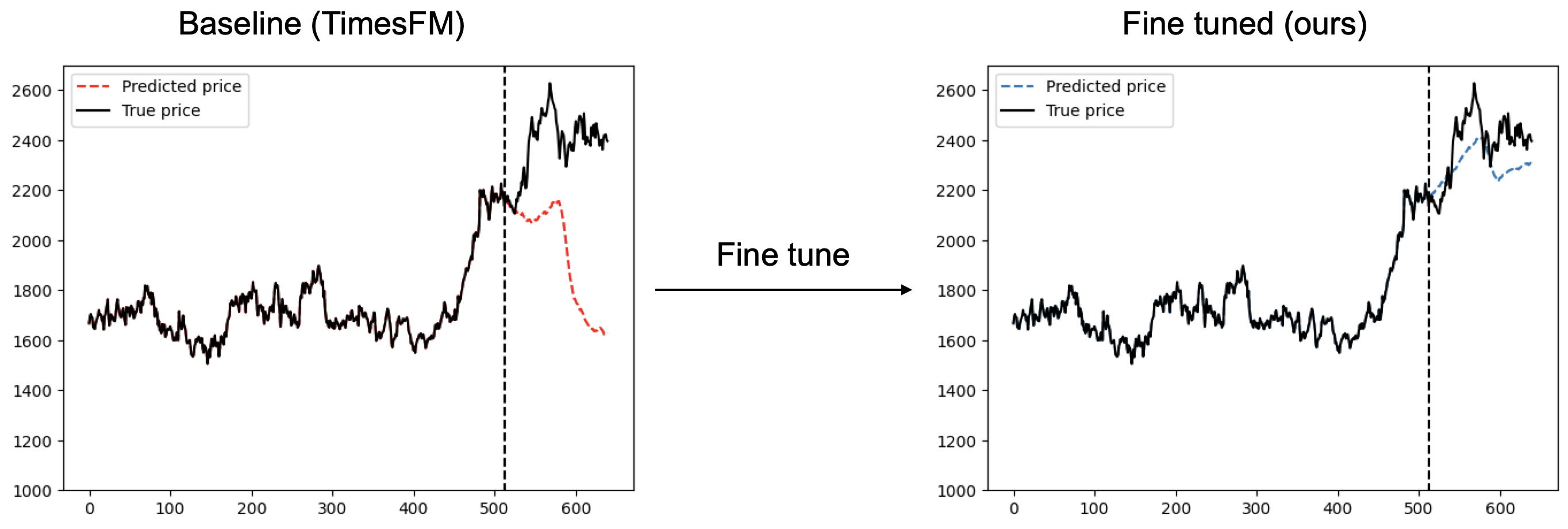}
    \caption{\label{figure1}We show that the baseline foundation time-series model TimesFM fails at the task of financial market price prediction. Fine-tuning TimesFM on financial data siginificantly improves its prediction accuracy.}
\end{figure}

In order to address this research question, we begin by evaluating the performance of TimesFM on the task of price prediction. We find the baseline TimesFM shows extremely undesirable performance. In particular, for the price trajectory prediction task shown in Figure 1, baseline TimesFM fails drastically. Nonetheless, we show that fine-tuning TimesFM through continual pre-training allows significant improvement over multiple standard benchmarks across several markets. We conduct a mock trading experiment with our fine-tuned TimesFM against multiple standard benchmarks to demonstrate its performance in the following markets: S\&P500 stocks, TOPIX500 stocks, forex market, cryptocurrencies.

We list our main contributions in this paper below:
\begin{enumerate}
    \item Dataset curation for fine-tuning on price data
    \item Fine-tuning TimesFM to give a \textbf{financial times series foundation model}
    \item Modification of training methods (loss and masking) to stabilize training on financial time series, specifically price data
    \item Testing of fine-tuned model by mock trading experiments to demonstrate that our fine-tuned model outperforms multiple standard benchmarks across a variety
    \item The code and model weights are published at \url{https://github.com/pfnet-research/timesfm_fin} for reproducibility of results and to accelerate further research
\end{enumerate}
The following paper discusses (in order): \textit{related work} (section \ref{relatedwork}) of transformers to time series predictions and a description of TimesFM, \textit{fine-tuning method} (section \ref{finetuning}) including our modifications, details about our \textit{experiments} (section \ref{experiments}) including training dataset, hyperparameter settings, evaluation metrics and mock trading. We follow this by the next section on \textit{results} (section \ref{results}), showing how fine-tuning TimesFM improves its performance over standard baseline a range of experiments and financial markets. We finally suggest some future work possibilities in \textit{discussions} (section \ref{discussion}).

\section{Related Work}
\label{relatedwork}
"Buy low, sell high" - one of the fundamental principles of the market that traders rely on to profit from their trades. But how do we know, what is 'low' and how much is 'high'? In essence, the problem reduces to accurate prediction of the price of a financial asset. 

Historically, quantitatively modelling prices have founded upon methods such as autoregressive \cite{box1970} combined with moving averages \cite{mckenzie1984} and conditional variances \cite{garch}, Kalman filtering \cite{kalman} and hidden Markov models \cite{hiddenmarkov} to name a few. Much of these models rely on well calibrated mathematical concepts to model the underlying dynamics of the market and fit a trend for the prices. Advancements in neural network architectures such as the RNN \cite{rnn} and LSTM \cite{lstm} have inspired price analysis \cite{lstmprice}. More recently, increased compute capabilities have shown that with large amounts of data and a sufficient model capacity to absorb the data, models can be trained to capture underlying trends and generalize as well, or even better than previous models. \cite{gpt2} A particular architecture which made this possible was the transformer.

Transformers, first introduced in \cite{Vaswani2017}, revolutionized natural language processing through large language models \cite{bert, gpt, llama, mixtral}, and computer vision fields of image classification \cite{vit}, video classification \cite{videovit} and image generation \cite{dit}. For our task of financial market price prediction, we examine some related works of utilizing transformers in time series predictions. \cite{timegpt1, timellm, patchtst, llmp, llmtime, timesfm}

Industry standard of using transformers for time series forecasting have made use mainly of LLMs \cite{llmp, llmtime, timellm, timegpt1, fingpt}. Recent works \cite{tan2024languagemodelsactuallyuseful, patchtst, llmtimeseriessurvey} question the necessity and relevance of LLMs in forecasting, and instead focus on training a time series foundation model with transformer building blocks specifically for the task of time series forecasting \cite{timesfm}, to which we turn our attention. 

TimesFM \cite{timesfm} is a 200-million parameter decoder-only model trained a time series datasets, containing 100-billion time points, with the task of next-value forecasting. We refer readers to the original paper \cite{timesfm}, but we provide a brief summary of this related work, including its methods and main findings in section \ref{finetuning}. 
  
Evaluation of TimesFM done on the Darts\cite{darts}, Monash \cite{monash} and Informer\cite{informer} benchmarks demonstrate strong performance of TimesFM in terms of the mean-average-error(MAE) metric in comparison to many previous SOTA methods, including N-BEATS \cite{nbeats}, LLM-Time \cite{llmtime}, ARIMA\cite{mckenzie1984} and PatchTST \cite{patchtst}.  
  
Our continual pre-training recipe is mostly aligned to that in TimesFM. More details are provided in Section \ref{finetuning}

\section{Financial Fine-tuning method}
\label{finetuning}
We first describe TimesFM, the existing model on which we base our fine-tuning. In TimesFM, input time series data is patched into $input\_patch\_len=l_i$ length patches, these are processed by stacked transformer layers from which an output block containing $output\_patch\_len=l_o$ time points are predicted. Mean-squared error (MSE) loss is computed on these $l_o$ points:
\begin{align}
\label{original_loss}
    Train\_Loss = \frac{1}{N}\sum_{j=1}^{N}MSE(\hat{y}_{l_i j+1:l_ij+l_o}, y_{l_ij+1:l_ij+l_o})
\end{align}
The authors typically set $l_i=32$ and $l_o=128$ and recommends $l_o>l_i$ to train the model in a decoder-only mode and also minimizes the number of autoregressive steps needed at inference time. Random masking is also applied to train the model going through all possible context lengths.  
  
During inference time, the model reads in the $l_o$ points it generates as input and repeatedly generates new time points until all have been autoregressively generated. Masking is not applied at inference time.  
  
Data used during pre-training of TimesFM mainly included Google trends, Wiki page views, and many other publicly available time series data sources. The authors also showed that a mix of synthetic data improved the performance of the model.  

In the remaining of this section, we introduce modifications to the original TimesFM for fine-tuning on financial data (specifically, price data). The method we employ is continual pre-training: restarting training from the pre-trained \href{https://huggingface.co/google/timesfm-1.0-200m}{weights} of TimesFM, continuing stochastic gradient descent on financial data. We restart training with a linear warmup to a learning rate of 5e-4 followed by a cosine decay. The specific training recipe is listed in Table \ref{table:hyperparameters} in section \ref{experiments}. Model architecture follows from the publicly available TimesFM \href{https://huggingface.co/google/timesfm-1.0-200m}{checkpoint}. We list two of our contributions for adapting TimesFM for continual pre-training on financial data.

\subsection{Loss}
The original MSE loss (Equation \ref{original_loss}) comes with its set of pitfalls when trained on price data: 
\begin{enumerate}
    \item Biases towards large scale values, e.g. a stock index with average values of USD1000 will receive much more weight in training than cryptocurrencies averaging BUSD0.0001 
    \item Instability due to market crash events. Especially when high price stocks experience a rapid crash of more than 99\% of its original value, instability in a single step results in NaN loss and failure of convergence. 
\end{enumerate}
In this section, we tackle these problems by describing a small modification to the loss computation. Namely, we apply a log transformation to the original time series, make predictions based on these transformed sequences. The MSE loss is then computed on these $\log$-ed sequences. What we do explicitly is 
\begin{align}
\label{log_transform}
    z \leftarrow \log(y)
\end{align}
where $z$ is used as the input to the model, then followed by 
\begin{align}
\label{log_loss}
    Train\_Loss = \frac{1}{N}\sum_{j=1}^{N}MSE(\hat{z}_{l_i j+1:l_i j+l_o}, z_{l_i j+1:l_i j+l_o})
\end{align}
For small changes in $y$, computing the MSE of $z=\log(y)$ is equivalent to computing the percentage MSE loss. But for large changes in $y$, the tapering of the $\log$ function results in a less than proportionate change in $z$, which in turn stabilizes training. 

\subsection{Masking}
We employ a similar masking strategy to that described in \cite{timesfm}, where we want to randomly sample the start and end points of the time series. This is done with the following method:  
  
For training efficiency, time series are broken up into sequences of length of at most $max\_context\_length + output\_length$.  We then randomly sample a random $t_{end}$ from $[min\_context\_length, max\_context\_length]$ then sample a random $t_{start}$ from $[0, t_{end}-min\_context\_length]$. The points between $[t_{start}, t_{end}]$ are then taken as input, where the model outputs the next $output\_len$ many points during training and loss is evaluated on those points.  
  
Typically, we set $min\_context\_len=128$ to ensure that the model is trained on meaningful (sufficiently long) examples. Our masking strategy fine tunes TimesFM to be able to predict any price data sequence of length from $min\_context\_length$ to $max\_context\_length$. These random masks change between batches and training steps, preventing overfitting by training the model to forecast from a variety of segments of the time series.   
  
Through the strategies describe in this section, we are able to complete fine tuning of TimesFM on 80M time points within 1 hour without any NaN loss. 

\section{Experiments}
\label{experiments}

In this section, we build upon our method described in section \ref{finetuning} and set up computational experiments to address our original research question: \textbf{can a foundation time series model perform well on price data in the financial markets?} We first begin by detailing the data and settings used to run our experiments: to build a fine-tuned TimesFM and compare it on several experiments against previous benchmarks (including the original TimesFM). These experiments, later explained more thoroughly, includes comparing the accuracy and F1-score of price prediction across various prediction horizons. From a financial perspective, we wish to understand the profitability of the model beyond evaluation metrics such as accuracy and F1-score which do not capture intricacies such as magnitudes of price movements, cost of trading among other considerations when deployed in the market. To that end, we propose a mock trading set up, devising an executable trading strategy based on our fine-tuned model, to compare foundation time series models (original and fine-tuned TimesFM) to a by-chance model and an AR1 model.

\subsection{Data}
Consisting of price time series in stocks, indices, foreign currencies and crypto currencies, data span granularities of hourly and daily. Main source used include Yahoo Finance and Binance, from which data is obtained using publicly available API endpoints. A detailed description of thecontinual pre-training data used can be found in Table \ref{table:dataset_summary}. Our dataset totals more than 100K time series and 90M time points.  
  
To avoid look-ahead bias, data from year 2023 onwards is reserved for testing. During the training process, we use a 75-25 split between train and validation dataset, randomly sampled from the same subset of time series ending before 1 Jan 2023.

\begin{table}[h!]
\centering
\caption{Summary of data used for fine-tuning.}
\label{table:dataset_summary}
\resizebox{\linewidth}{!}{
\begin{tabular}{|l|l|r|r|}
\hline
\textbf{Dataset} & \textbf{Granularity} & \textbf{\# Time Series} & \textbf{\# Time Points} \\ 
\hline
Topix500 stocks & Daily & 3513 & 2,248,320 \\ 
\hline
S\&P500 stocks & Daily & 3173 & 2,030,720 \\ 
\hline
Currencies & Daily & 1092 & 698,880 \\ 
\hline
Japan Investment Trusts & Daily & 6698 & 4,286,720 \\ 
\hline
Commodities & Daily & 29 & 18,560 \\ 
\hline
Stock Indices & Daily & 216 & 138,240 \\ 
\hline
Stock Indices & Hourly & 847 & 542,080 \\ 
\hline
Stock prices & Hourly & 31,756 & 20,323,840 \\ 
\hline
Cryptocurrencies & Daily & 1680 & 1,075,200 \\ 
\hline
Cryptocurrencies & Hourly & 79,153 & 50,657,920 \\ 
\hline
\end{tabular}
}
\end{table}
  
As opposed to the original TimesFM, we do not use any synthetic data in training and do not conduct any reweighting to sample each granularity evenly. We acknowledge the possibilities of future work in this area where some suggestions are offered in Section \ref{discussion}. Nonetheless, we observe that while the training process included more of hourly granularity data, the model shows better performance over longer prediction horizons, as demonstrated in Section \ref{results}.

\subsection{Hyperparameters}
Table \ref{table:hyperparameters} list out the settings used for fine-tuning TimesFM. Notably, we use SGD with linear warmup and cosine decay with a peak learning rate of 5e-4. 
\begin{table}[h!]
\centering
\caption{Hyperparameters and architecture settings.}
\label{table:hyperparameters}
\begin{tabular}{|l|l|}
\hline
\textbf{Hyperparameter/Architecture} & \textbf{Setting}   \\
\hline
Optimizer                            & SGD                \\
\hline
Linear warmup epochs                 & 25                 \\
\hline
Total epochs                         & 100                \\
\hline
Peak learning rate                   & 5e-4               \\
\hline
Momentum                             & 0.9                \\
\hline
Gradient clip (max norm)             & 1.0                \\
\hline
Batch size                           & 1024               \\
\hline
Max context length                   & 512                \\
\hline
Min context length                   & 128                \\
\hline
Input length                         & 32                 \\
\hline
Output length                        & 128                \\
\hline
Layers                               & 20                 \\
\hline
Hidden dimensions                    & 1280               \\
\hline
\end{tabular}
\end{table}

Following the training recipe listed out in Table \ref{table:hyperparameters} and using the data in \ref{table:dataset_summary}, we are able to complete training on 8 V100s within 1 hour without any NaN loss. Training curves are shown in Figure \ref{loss_curves}. 

\subsection{Testing}
\label{testing}
To explore whether fine-tuning TimesFM does indeed lead to a performance gain when deployed in financial market, we run several experiments detailed in this section. The first metric we compare is the accuracy of price predictions, over various prediction horizons (equivalent to holding period of the asset). This is done on the test set (data from 2023 onwards, not used in training and validation). We also introduce a more robust metric: Macro F1-score, allowing more fair comparisons of the models even under class-imbalanced situations. Lastly, we conduct mock trading in various markets: S\&P500 stocks, TOPIX500 stocks, currencies, cryptocurrencies, to verify that performance in accuracy and macro F1 is translated into Profit and Loss (PnL). 

\subsubsection{Metric: Accuracy}
\label{accuracy}
Recall that at training time, the model is given $input\_length<=max\_context\_length=512$ data points (with random masking) and tasked to always predict the next $output\_len=128$ many points. Loss is evaluated on these $output\_len$ many points. 
  
At inference time, the model is consistently given $context\_length=c$ many points (without masking, where $c<=512$) and tasked to predict the following points. However, we might wish to generate an arbitrary number of future points, not necessarily 128.  
  
At each step, the model predicts the next $h$ many points. For this single step, accuracy is evaluated on the last output point $y_{c+h}$, where the model is tasked to classify whether the price moves up or down. At the next step, the model then reads in the next ground truth $h$ points, and computes the accuracy again with $\hat{y}_{c+2h}$. Accuracy calculation is based on this classification over every inference step, i.e.

\begin{align}
\label{accuracy_eqn}
    Accuracy(\hat{y}_{c+kh}, y_{c+kh} | y_{1:c+(k-1)h})
\end{align}
for all $1 \leq k \leq K$ such that $Kh$ is longer than the desired total prediction horizon $H$. In our experiments, we fix $H=128$ and vary $h \in \{2, 4, 8, \dots, 128\}$. We note that the model is only capable of processing the last $max\_context\_length=512$ points in $y_{1:c+(k-1)h}$.

Results comparing our fine-tuned TimesFM against the baseline TimesFM are shown in Figure \ref{accuracy_graph}. 

\subsubsection{Metric: F1 score}
\label{f1}
Accuracy scores may not always be a reasonable metric of interest. As an example, on a biased dataset with 90\% positive samples and 10\% negative. a model classifying every sample as positive will attain a 90\% accuracy. Nonetheless, we should ask the question: did the model truly learn the underlying distribution.

An alternative to accuracy is the F1-score, taken as the harmonic mean of precision and recall. However, the F1-score also suffers from problems with class imbalance \cite{f1imbalance}, which points us to adopt the Macro F1-score \cite{macrof1}, which is computed as the arithmetic mean of F1-scores taking each class as the 'positive' class. Specifically, in the example above we obtain a Macro F1-score of 0.474, showing that the model fails to learn the 10\% negative rate well. 

\subsubsection{Mock trading}
\label{mock_trading}
Here, we develop trading strategies based on fine-tuned TimesFM and analyze our profits from our trades.

We outline below the first trading strategy, which we shall term \textit{basic strategy}. The trader begins by selecting a holding period, denoted as \( h \) (where \( h = \text{horizon\_len} \)). Additionally, we define the context length as \( c \) (where \( c = 512 \), using maximum context length throughout). $h$ and $c$ are analogous to that in Equation \ref{accuracy_eqn}, since the trades are made based on the prediction $h$-steps ahead. 
  
After trading day \( i \), the trader inputs the time series \( P_{i-c-1:i} = \{P_{i-c-1}, P_{i-c}, \dots, P_i\} \) into a model to obtain a prediction for \( P_{i+1:i+h} \). The trader places a buy or sell order on day \( i+1 \) and \( i+h \) based on the following conditions:

\begin{itemize}
    \item If \( P_{i+h} > P_{i+1} \), place a buy order on day \( i+1 \) and a sell order on day \( i+h \).
    \item If \( P_{i+h} < P_{i+1} \), place a sell order on day \( i+1 \) and a buy order on day \( i+h \).
\end{itemize}
This strategy is repeated for all trading days \( i \). 
  
If the trading basket contains a total of \( T \) assets, all orders placed will be worth \( \frac{1}{(h-1)T} \) each. This ensures that the \( \ell_1 \)-norm of the orders placed on a given day does not exceed \( \frac{1}{h-1} \), and over the entire holding period does not exceed 1 (which represents the total capital). We limit the norm of our orders to ensure that, even in the extreme case where every order during the holding period is a ‘buy’, sufficient capital is available to place all orders. As an example, our initial budget is \$1, and the holding period $h=100$. If for all $0 \leq i \leq 99$, we predict $P_{i+h} > P_{i+1}$, then for we will place buy orders for all $0 \leq i \leq 99$. Hence the maximum we can trade in one day will be limited to $\$\frac{1}{99}$. \\
An alternative strategy we compare is the \textit{market neutral strategy}. A concern of the \textit{basic strategy} is how our portfolio positions depend on the overall market bias, and will be affected by the total market movement. This is also illustrated in Figure \ref{sp500_ours_simple}. To construct a market neutral strategy such that our returns are independent of the overall market movement, the mean position for each day is subtracted. As an example, if a trade a basket of three stocks \textit{A, B, C}, and our \textit{basic strategy} positions were $-1/3, 1/3, 1/3$ respectively, then our \textit{market neutral strategy} positions will be $-4/9, 2/9, 2/9$, i.e. this mean subtraction is done on top of the \textit{basic strategy}. Hence, instead of limiting our dailu budget, we limit our daily exposure to $1/(h-1)$. The results for this is shown in Figure \ref{sp500_ours_simple}.

\section{Results}
\label{results}
This section gives a comprehensive analysis of the fine-tuned TimesFM results following the experiments described in section \ref{experiments} and its comparison to the original version as well as some popular past benchmarks.
\subsection{Training results: Loss curves}
\begin{figure}[h!]
    \centering
    \includegraphics[width=0.45\linewidth]{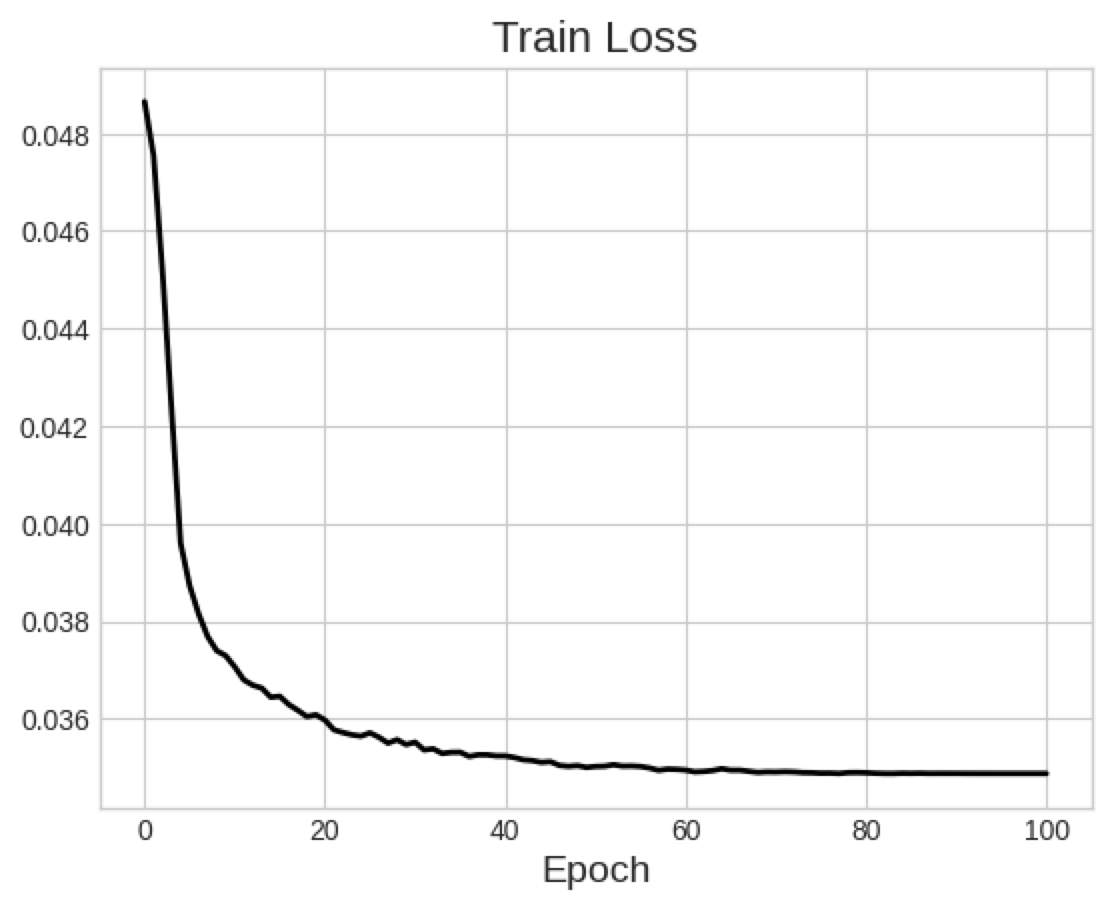}
    \includegraphics[width=0.45\linewidth]{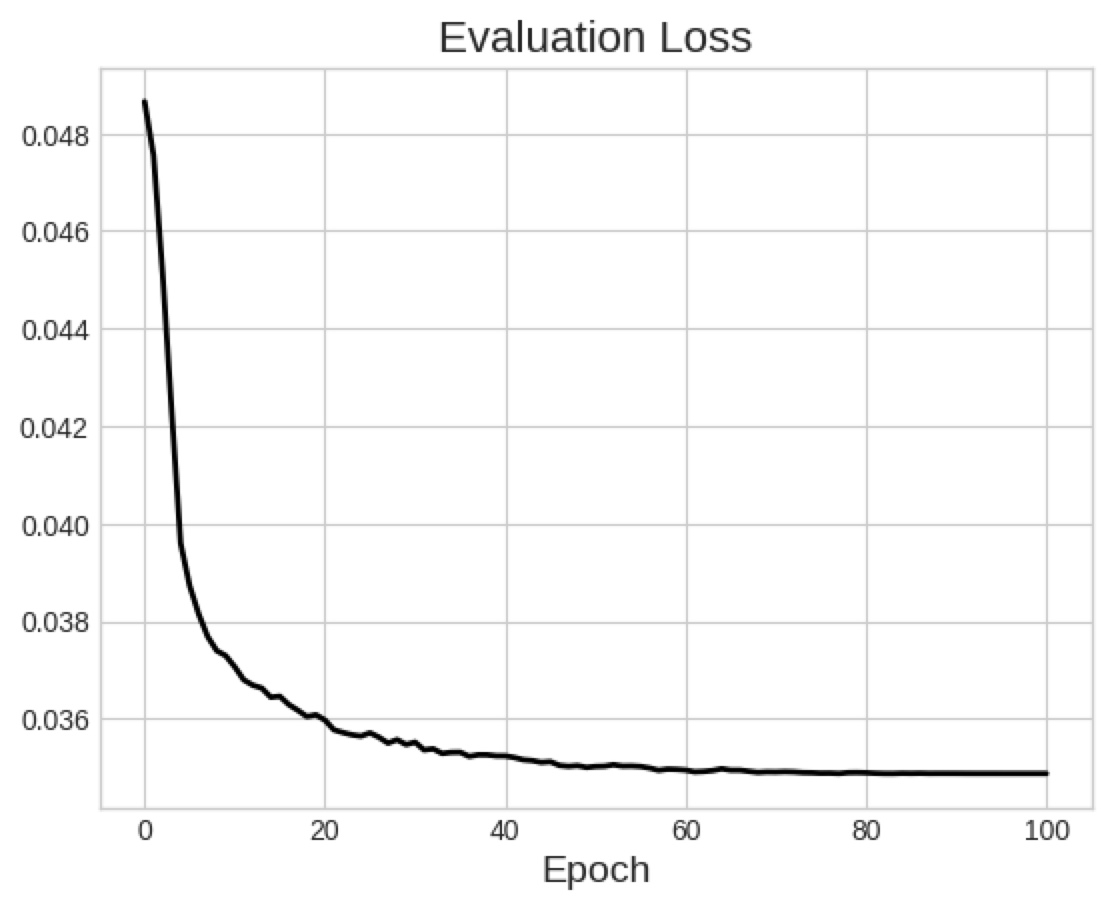}
    \caption{\label{loss_curves}Training and Validation loss curves for fine tuning TimesFM following the recipe in Table \ref{table:hyperparameters}. Training generally asymptotes at around 70\% of the original loss value.}
\end{figure}
As shown in Figure \ref{loss_curves}, training usually asymptotes at around 70\% of the original loss value. Noise in training is present due to the random masking augmentation described in the previous section. Note that extending training past 100 epochs, or using a larger learning rate gives preliminary signs of overfitting. We recommend, for future work, the use of larger training set, stronger data augmentation or early stopping for better generalization. 
  
We note that the loss values in Figure \ref{loss_curves} display the loss after performing the $\log$ transformation, and does not immediately translate into the same performance when computing MSE loss evaluated on the original samples. We verified that MSE loss also certainly decreases.  
  
While this demonstrates learning capabilities of the model, MSE loss can be easily reduced by many methods. Due to the similarities between train and validation sets (both drawn from data before 2023), overall market trends and high correlation between prices can incentivize a model to learn by memorization of seen patterns. 
  
In the following sections, we evaluate the performance of this model on the test set: data from 2023 onwards.

\subsection{Metric: Accuracy}
\label{accuracy_results}
\begin{figure}[h!]
    \centering
    \includegraphics[width=0.8\linewidth]{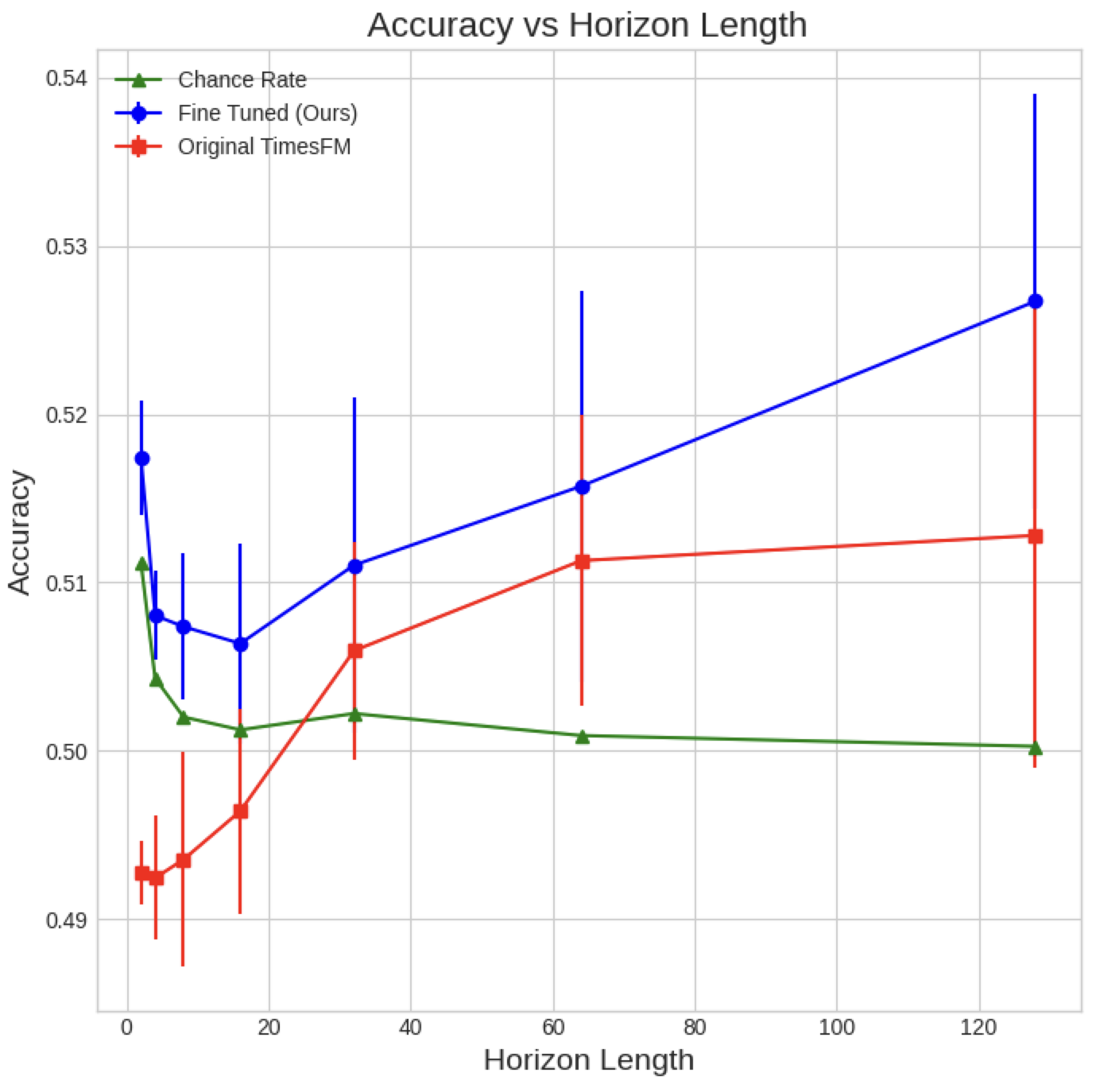}
    \caption{\label{accuracy_graph}Accuracy score of fine-tuned versus original TimesFM and a chance-rate model, when evaluated on the full test set.}
\end{figure}
We observe, in Figure \ref{accuracy_graph}, that through fine-tuning, we are able to see consistently better performance over the vanilla pre-trained TimesFM across horizon lengths from 2 to 128. As a benchmark, we provide the chance rate, calculated as the accuracy obtained by a random model. For example, if 53\% of the price changes in the test set is up, the random model guesses up 53\% of the time, and down 47\% of the time. Our fine-tuned TimesFM is also able to outperform this benchmark, providing statistical confidence that the improvements we see are not due to random chance.

As an answer to our original research question, the following conclusions can be drawn from Figure \ref{accuracy_graph},
\begin{enumerate}
    \item Original TimesFM underperforms random chance on 4 out of 7 of the prediction horizons, suggesting that TimesFM cannot be used it its original state for price prediction
    \item Fine-tuned TimesFM outperforms original TimesFM on all prediction horizons, showing how fine-tuning on financial data significantly improves performance.
    \item Fine-tuned TimesFM outperforms random chance on all prediction horizon, hinting at statistically significant performance.  
\end{enumerate}

\subsection{Metric: F1-score}
\label{f1_results}
\begin{figure}[h!]
    \centering
    \includegraphics[width=0.8\linewidth]{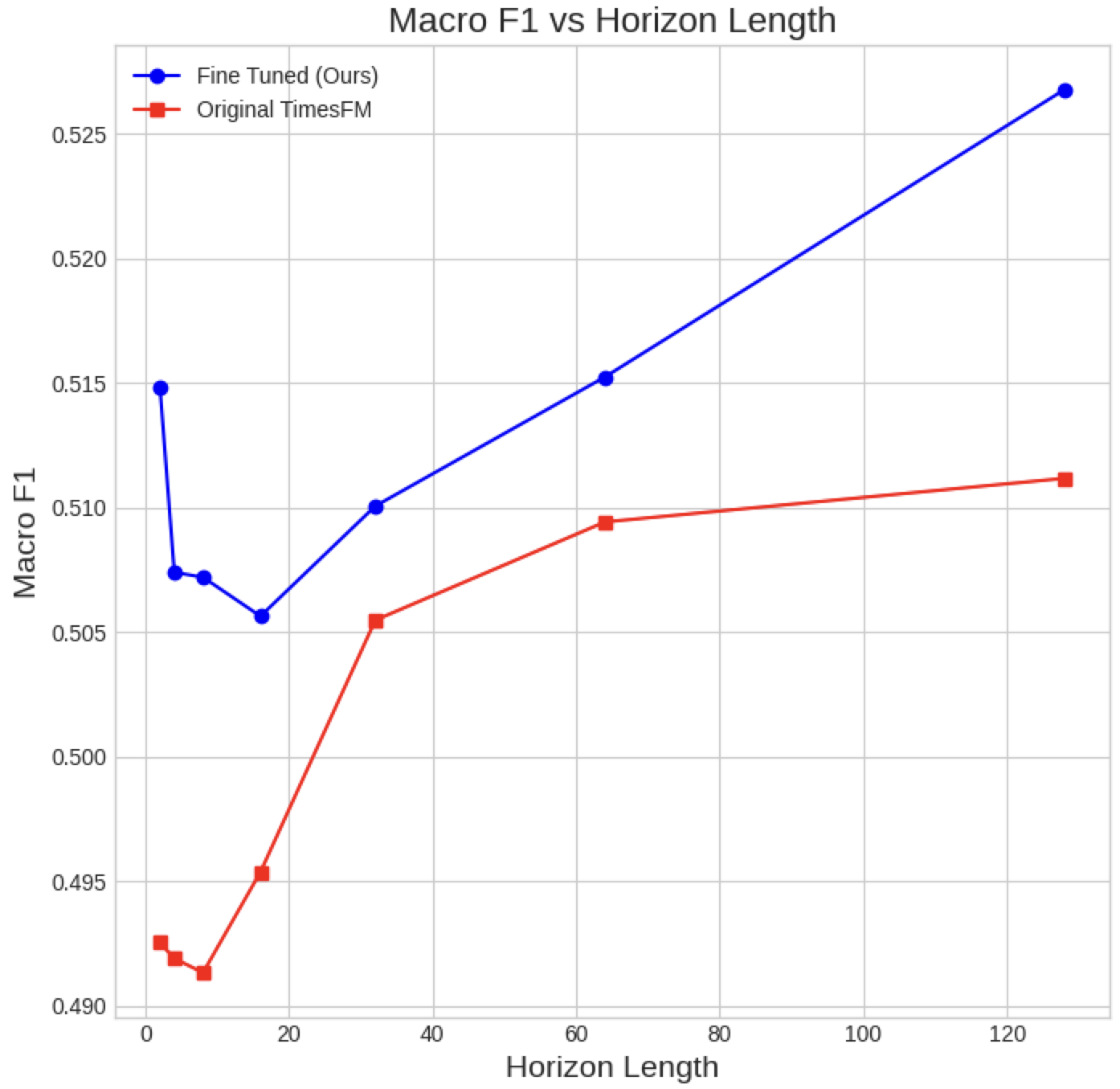}
    \caption{\label{f1_graph}Macro F1 score of fine-tuned versus original TimesFM, when evaluated on the full test set.}
\end{figure}
A more rigorous of the model performance with the Macro F1-score shown in Figure \ref{f1_graph} show identical trends to Figure \ref{accuracy_graph}, where the fine-tuned TimesFM consistently outperforms random chance and the baseline model. This suggests a conclusive answer to our original research question: \textbf{foundation time series models can perform well on price data in the financial markets after fine-tuning.}

\subsection{Mock Trading}
\label{mock_trading_results}
\begin{figure}[h!]
    \centering
    \includegraphics[width=\linewidth]{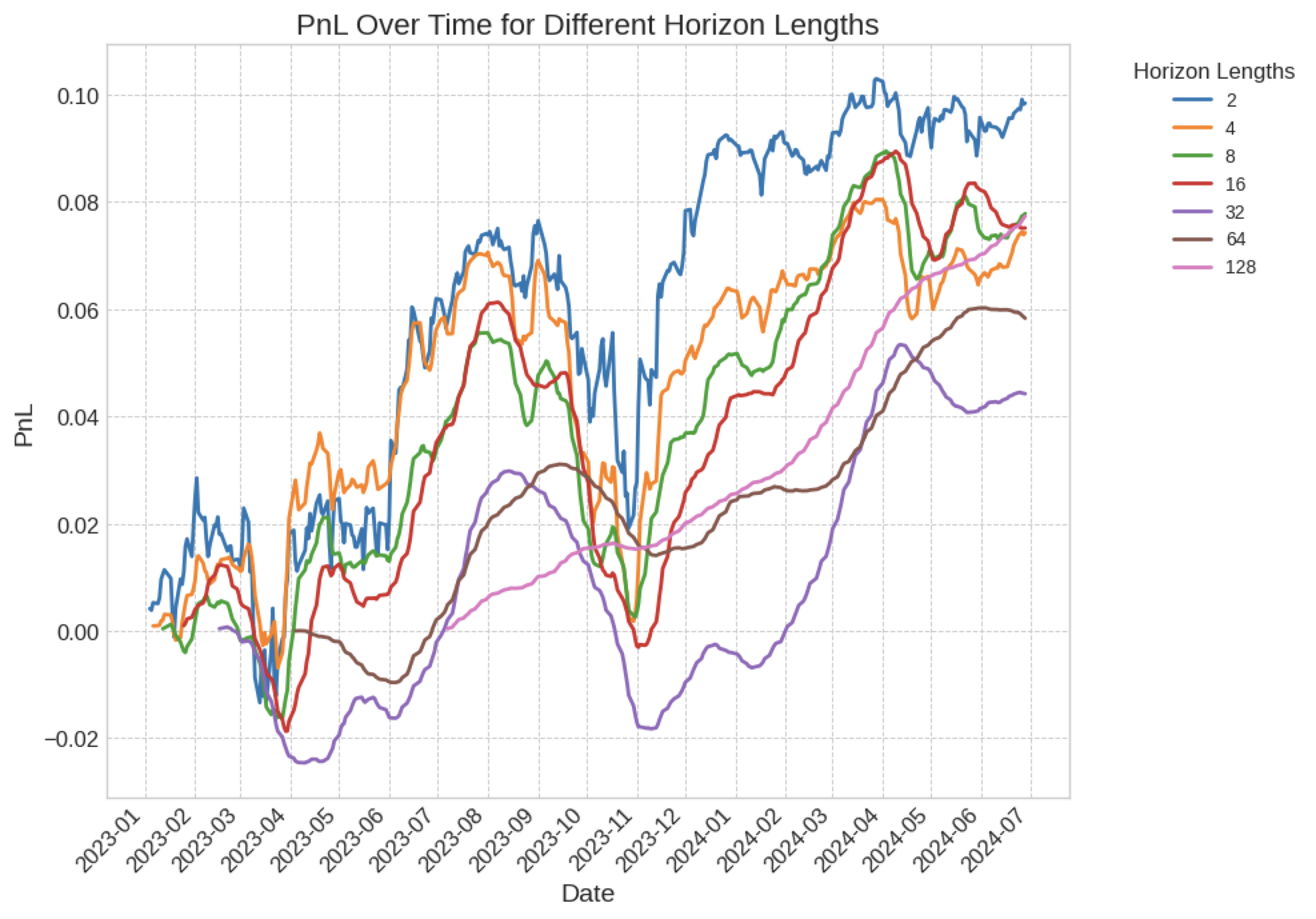}
    \caption{\label{sp500_ours_simple}Realized PnL using fine-tuned TimesFM traded on S\&P500 stocks using the basic strategy, assuming no trading costs.}
\end{figure}

In the previous section, we have demonstrated that our fine-tuned model is able to outperform standard benchmarks (original TimesFM and a chance-rate model) in price prediction tasks by a reliable and significant margin. Using the strategy outlined in section \ref{mock_trading}, we conduct mock trading of the fine-tuned TimesFM on daily data from the S\&P 500 index starting from January 1, 2023. The returns generated from executing this strategy (in a zero cost setting) are presented in Figure \ref{sp500_ours_simple}.

Using the basic strategy, we see consistently positive gains over each horizon length at the end of the trading period. Note that using a horizon length of $H$, the first returns will only be realized on day $H$, so the start points for each horizon length is different. 

\begin{figure}[h!]
    \centering
    \includegraphics[width=\linewidth]{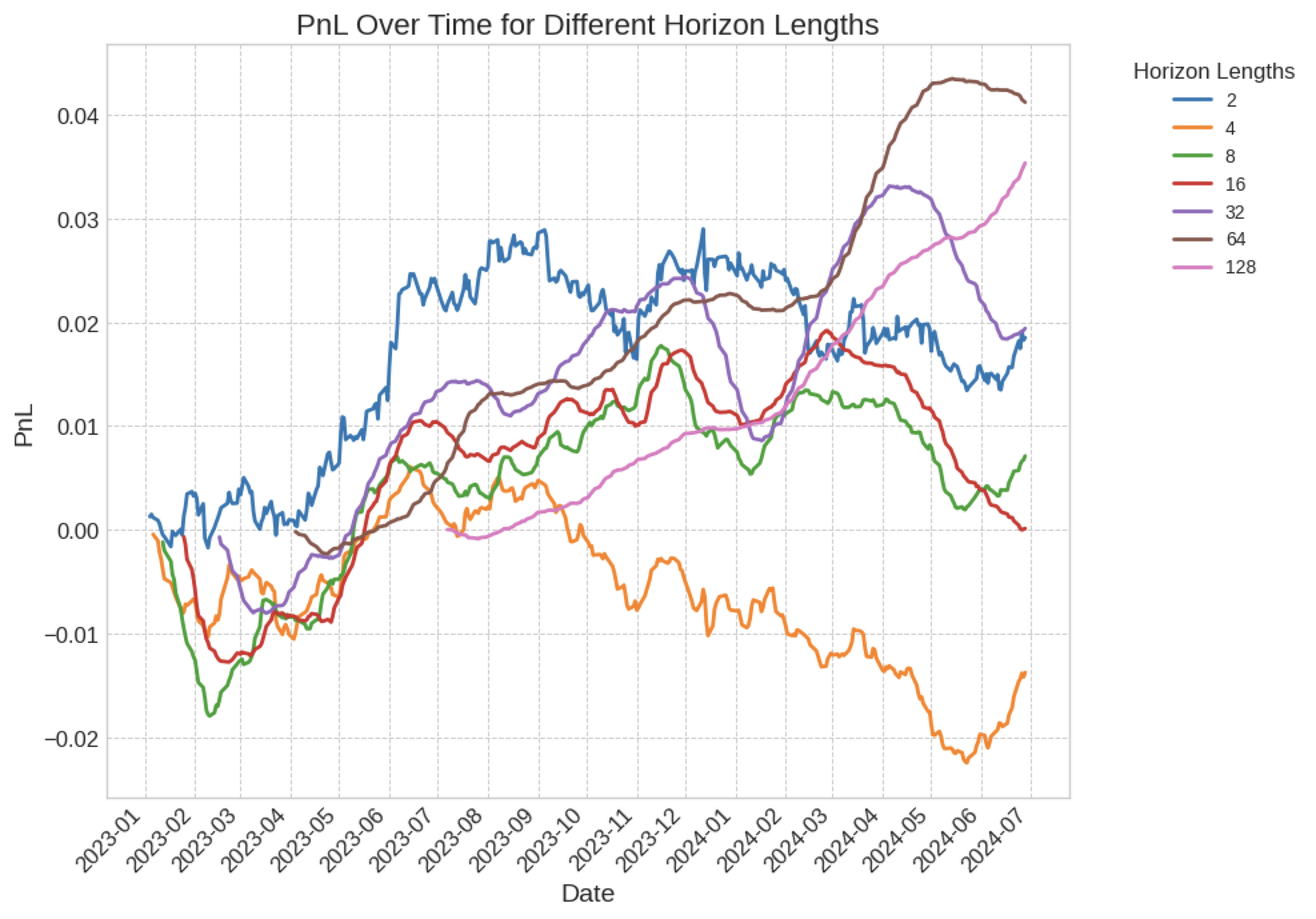}
    \caption{\label{sp500_ours_neutral} Realized PnL using fine-tuned TimesFM traded on S\&P500 stocks using the market neutral strategy, assuming no trading costs. }
\end{figure}

While we are able to observe maximum returns of up to 10\% (using horizon length of 2), this basic strategy is highly volatile, due to its dependence on the overall market movement. In contrast, results in Figure \ref{sp500_ours_neutral} show that the market neutral strategy is effective in reducing the overall volatility while ensuring positive returns for majority of horizon lengths.  

Furthermore, we provide additionally useful metrics for evaluating the performance of the market neutral strategy. 

\begin{table}[h!]
\caption{Performance Metrics by Horizon}
\centering
\resizebox{\linewidth}{!}{%
\begin{tabular}{|c|c|c|c|c|c|}
\hline
\textbf{Horizon} & \textbf{Ann Sharpe} & \textbf{Max Drawdown} & \textbf{Ann Returns} & \textbf{Ann Volatility} & \textbf{Neutral Cost (\%)} \\ \hline
2   & 0.516  & -0.015  & 0.013  & 0.024  & 0.003  \\ \hline
4   & -0.483 & -0.028  & -0.009 & 0.019  & -0.006 \\ \hline
8   & 0.227  & -0.017  & 0.005  & 0.022  & 0.007  \\ \hline
16  & 0.003  & -0.019  & 0.000  & 0.024  & 0.000  \\ \hline
32  & 0.420  & -0.015  & 0.014  & 0.034  & 0.080  \\ \hline
64  & 1.285  & -0.002  & 0.033  & 0.026  & 0.347  \\ \hline
128 & 1.679  & -0.001  & 0.036  & 0.021  & 0.600  \\ \hline
\end{tabular}%
}
\end{table}

In general, we observe the more desirable performance for larger horizon lengths, with $H=128$ achieving annual returns of 3.6\% and an annual sharpe of 1.68. Neutral cost, defined as the cost basis to zero out returns at the end of the trading period, increases with larger horizon lengths as we average the price movements out over slower moving strategies. Again, the largest horizon length of 128 can be traded up to a cost of 0.60\%. 

Next, we focus on a horizon length of 128 and compare our proposed model against several others. We focus on the market neutral strategy, comparing our proposed fine-tuned TimesFM against the original TimesFM, as well as a random model and an AR(1) model \cite{box1970}.

For the construction of a random model, we proceed according to the chance rate calculation presented in Figure \ref{accuracy_graph}. To recap, this model is created by first calculating the ratio up:down within the whole dataset, then at each day $i$, guess the sign of $P_{i + H} - P_{i + 1}$ according to this ratio.

The AR(1) model is an autoregressive model fitted with only the single-difference lagged term. To implement this, on each time series (in the training period) we fit an AR(1) model to obtain the coefficients, then predict on the test dates. Then, mean subtraction is done to turn this into a market neutral strategy.

\begin{figure}[h!]
    \centering
    \includegraphics[width=\linewidth]{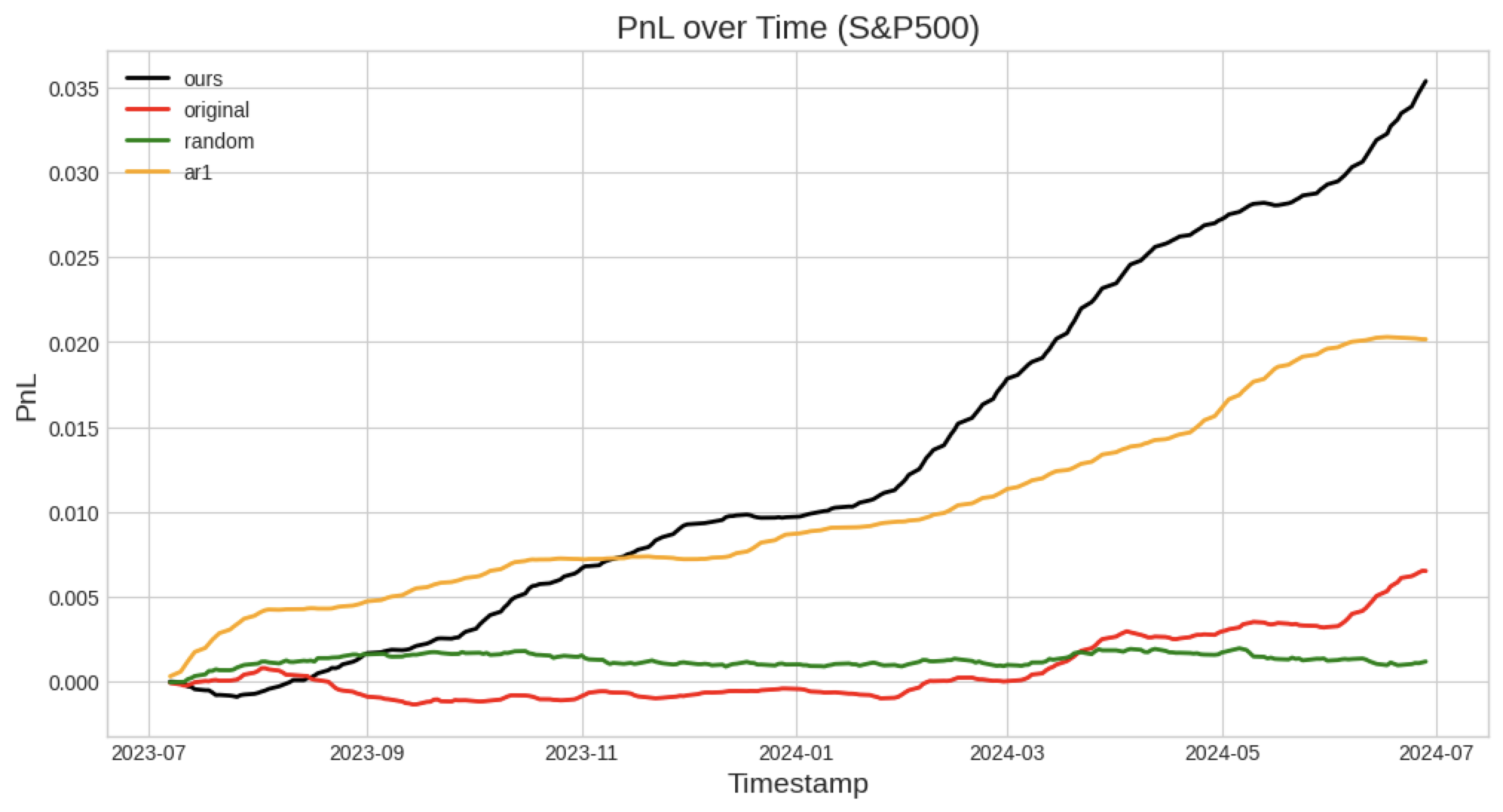}
    \caption{\label{sp500_comparison} Realized PnL comparison between various models traded on S\&P500 stocks using the market neutral strategy in a cost-free setting.}
\end{figure}

Results shown in Figure \ref{sp500_comparison} illustrate the stronger performance of our fine-tuned TimesFM over other models. The random model performs poorly in a market neutral setting, while the AR1 model also shows lower returns than fine-tuned TimesFM. 

Rigorous testing results across different markets, evaluated on sharpe ratio and neutral cost, are shown in Tables \ref{tab:performance_metrics_sharpe} and \ref{tab:performance_metrics_cost} respectively. 

\begin{table}[h!]
\caption{Comparison of Sharpe Ratio across models/markets}
\centering
\resizebox{\linewidth}{!}{
\begin{tabular}{lcccc}
\hline
               & \textbf{Ours} & \textbf{Original TimesFM} & \textbf{Random} & \textbf{AR1} \\
\hline
\textbf{S\&P500}      & 1.68 & 0.42  & 0.03  & 1.58  \\
\textbf{TOPIX500}     & 1.06 & -1.75 & 0.11  & -0.82 \\
\textbf{Currencies}   & 0.25 & -0.04 & -0.03 & 0.88  \\
\textbf{Crypto Daily} & 0.26 & -0.03 & 0.01  & 0.17  \\
\hline
\end{tabular}
}
\label{tab:performance_metrics_sharpe}
\end{table}

Our model outperforms the original TimesFM on all benchmarks, and the random model cannot make any reliable predictions under a market neutral situation.

However, performance of our model on currencies and crypto is left to be desired. Significantly underperforming the AR1 model. Nonetheless, our fine-tuned TimesFM is still the only model to achieve positive returns on every market.

\begin{table}[h!]
\caption{Comparison of Neutral Cost across models/markets}
\centering
\resizebox{\linewidth}{!}{
\begin{tabular}{lcccc}
\hline
               & \textbf{Ours} & \textbf{Original TimesFM} & \textbf{Random} & \textbf{AR1} \\
\hline
\textbf{S\&P500}      & 0.60\%  & 0.11\%   & -0.008\% & 0.34\%  \\
\textbf{TOPIX500}     & 0.14\%  & -0.24\%  & 0.02\%   & -0.18\% \\
\textbf{Currencies}   & 0.08\%  & -0.017\% & -0.008\% & 0.27\%  \\
\textbf{Crypto Daily} & 0.44\%  & -0.07\%  & 0.010\%  & 0.88\%  \\
\hline
\end{tabular}
}
\label{tab:performance_metrics_cost}
\end{table}

\section{Discussion}
\label{discussion}
In section \ref{results}, we have rigorously shown the viability of fine-tuning a foundation time series model (TimesFM) for the task of price prediction on financial markets. Nonetheless, our results raise several questions and motivations for future work. 

In preparation of the fine-tuning data, we used a mix of data from various markets and granularities. However, majority of training data was dominated by hourly cryptocurrency and stock data, which may result in biases during training towards a specific granularity or market. One could potentially upsample the underrepresented granularity or market data to balance the dataset, as was done in TimesFM, but keeping in mind the repitition of data that might deteriorate model performance. 

In TimesFM, it was also shown that including synthetic data in training, specifically time series given by simple mathematical functions, improves model performance even when evaluated on real-world time seris information. In other modalities, synthetic data has also shown to benefit model performance \cite{synthetic_image, synthetic_video} and the authors question to what extent synthetic data can benefit a time series model for financial price prediction here. 

During the training process, several decisions were empirically made about the loss function and the masking scheme to tweak TimesFM for fine-tuning on financial data. Another potential loss function can be to compute $\log(MSE)$ instead of $MSE(\log)$ and our preliminary observations show that they give rather similar results. Out of the scope of this paper, but more recently supported by the TimesFM authors, is training with quantile loss where the model also outputs confidence scores and quantiles during inference time.

While we chose to fine-tune with continual pre-training, this is one of the slowest methods of fine-tuning, which was only feasible in this situation due to the rather limited size of our dataset. Such a fine-tuning method also increases the magnitude of the change in the model weights compared to before fine-tuning. Other alternatives such as freezing model weights, LoRA \cite{lora} can help the model make smaller updates during fine-tuning while still achieving desired performance. 

In our evaluation experiments, we have also seen that the original TimesFM significantly underperforms even the most basic AR1 model, giving statistically insignificant and even negative returns on most of the markets. While this shows the benefit of fine-tuning TimesFM, we question where the performance bottleneck in the original TimesFM lies. The authors hypothesize the irregularities of price data compared to the regular time series data that TimesFM is trained on to be a primary factor causing TimesFM to be unable to capture the underlying market dynamics. Additional experiments of original TimesFM on a wider range of data complexities, granularities, trading periods can help to elucidate the differences. A helpful comparison would be the difference between the original and fine-tuned weights of TimesFM. 

We observed that accuracy in price prediction task improves, what happens to the performance on generic time series forecasting? Specifically within language models, fine-tuning can deteriorate generalization performance by destroying pretrained features \cite{kumar2022finetuning}. Specifically for a fine-tuning dataset like ours, containing price data with little to no correlation with standard time series data, fine-tuning has the potential of worsening performance on general benchmarks. A next step to this would be to evaluate the model through MAE (mean average error) scores on benchmarks like Darts\cite{darts}, Monash \cite{monash} and Informer\cite{informer} as was used in TimesFM.

While fine-tuning improves TimesFM over its baseline, we are unable to ascertain consistently better performance over just a simple AR1 model. How should we improve the fine-tuning to surpass AR1? How would its performance compare to other autoregressive models? Possible directions include crafting a better fine-tuning dataset with balance over different granularities \cite{timesfm}, or adjusting the loss function to perform probabilities forecasting through quantile loss. 

This also raises the question of what exactly is the model learning? Comparisons in Tables \ref{tab:performance_metrics_cost} and  \ref{tab:performance_metrics_sharpe} show that it is not entirely just based off a single autoregressive term, else our metrics would have strong similarity to AR1 across all markets. Computing the correlation in the predicted prices between TimesFM and AR1, as well as with other autoregressive models, or probing the internal activations with linear probing techniques \cite{probing}, can help explore what kind of momentum (or some other) strategies a large time series model is learning.

\section{Conclusion}
In this paper, we have fine-tuned a time series foundation model \cite{timesfm} for usage on financial data. By evaluating the loss and accuracy of the model, we see that fine-tuning allows to achieve significantly superior results using the large capacity of TimesFM, outperforming traditional models.

We tested this model by constructing a trading strategy that places buy/sell trades according to the predictions of the model. Through thorough evaluation, we found that a market neutral strategy with a long horizon gives consistently better performance over traditional models, with a sharpe ratio up to 1.68 when traded on S\&P500.

We publish our \href{https://github.com/pfnet-research/timesfm_fin}{code} and \href{https://huggingface.co/pfnet/timesfm-1.0-200m-fin}{model weights} for reproducibility of results, and hope it inspires future research in this direction.

\bibliographystyle{IEEEtran}
\bibliography{cite}

\end{document}